\documentclass[letterpaper]{article} % DO NOT CHANGE THIS
\usepackage{aaai24}  % DO NOT CHANGE THIS
\usepackage{times}  % DO NOT CHANGE THIS
\usepackage{helvet}  % DO NOT CHANGE THIS
\usepackage{courier}  % DO NOT CHANGE THIS
\usepackage[hyphens]{url}  % DO NOT CHANGE THIS
\usepackage{graphicx} % DO NOT CHANGE THIS
\usepackage{natbib}  % DO NOT CHANGE THIS AND DO NOT ADD ANY OPTIONS TO IT
\usepackage{caption} % DO NOT CHANGE THIS AND DO NOT ADD ANY OPTIONS TO IT
\usepackage{algorithm}
\usepackage{newfloat}
\usepackage{listings}
\DeclareCaptionStyle{ruled}{labelfont=normalfont,labelsep=colon,strut=off} % DO NOT CHANGE THIS
\floatstyle{ruled}
\newfloat{listing}{tb}{lst}{}
\floatname{listing}{Listing}
\usepackage{booktabs} % For formal tables
\usepackage{enumitem}
\usepackage{multirow}
\usepackage{amsmath}
\usepackage{cleveref}
\usepackage{marginnote}

\usepackage[dvipsnames]{xcolor}
\usepackage{soul}
\definecolor{navy}{rgb}{0.0, 0.0, 0.0}
\definecolor{ruby}{rgb}{0.0, 0.0, 0.0}

\newcommand{\sw}[1]{{\color{black}{#1}}}

\usepackage{xspace} %% smart spacing

\usepackage{array} % for better arrays (eg matrices) in maths
\usepackage{algpseudocode}
\title{How to Train Your YouTube Recommender to Avoid Unwanted Videos}

\author {
Alexander Liu, 
Siqi Wu, 
Paul Resnick
}

\affiliations {
University of Michigan, Ann Arbor\\
\{avliu, siqiwu, presnick\}@umich.edu\\
}
\begin{document}
\maketitle

% ====================================================
\begin{abstract}

YouTube provides features for users to indicate disinterest when presented with unwanted recommendations, such as the ``Not interested'' and ``Don't recommend channel'' buttons. These buttons purportedly allow the user to correct ``mistakes'' made by the recommendation system. Yet, relatively little is known about the empirical efficacy of these buttons. Neither is much known about users' awareness of and confidence in them. To address these gaps, we simulated YouTube users with sock puppet agents. Each agent first executed a ``stain phase'', where it watched many videos of an assigned topic; it then executed a ``scrub phase'', where it tried to remove recommendations from the assigned topic. Each agent repeatedly applied a single scrubbing strategy, either indicating disinterest in one of the videos visited in the stain phase (disliking it or deleting it from the watch history), or indicating disinterest in a video recommended on the homepage (clicking the ``not interested'' or ``don't recommend channel'' button or opening the video and clicking the dislike button). We found that the stain phase significantly increased the fraction of the recommended videos dedicated to the assigned topic on the user's homepage. For the scrub phase, using the ``Not interested'' button worked best, significantly reducing such recommendations in all topics tested, on average removing 88\% of them. Neither the stain phase nor the scrub phase, however, had much effect on videopage recommendations (those given to users while they watch a video). We also ran a survey ($N$ = 300) asking adult YouTube users in the US whether they were aware of and used these buttons before, as well as how effective they found these buttons to be. We found that 44\% of participants were not aware that the ``Not interested'' button existed. Those who were aware of it often used it to remove unwanted recommendations (82.8\%) and found it to be modestly effective (3.42 out of 5).

\end{abstract}
% ====================================================

% ====================================================
\section{Introduction}
\label{sec:intro}

YouTube is the world's largest long-form video sharing platform, with users watching a billion hours of YouTube's content every day \cite{goodrow_you_2017}. In recent years, the YouTube recommendation algorithm has come under increased scrutiny for its role in promoting conspiracy theories \cite{tomlein_audit_2021}, radical content \cite{hosseinmardi_examining_2021,hosseinmardi2024causally}, and Alt-Right ideology \cite{lewis_alternative_2018,ribeiro_auditing_2020}. Studies have found that watching such content can lead to its continual and sometimes increased promotion \cite{ledwich_algorithmic_2019,hussein_measuring_2020}. Besides societally-harmful information, YouTube has also been known to make unwanted recommendations to individuals, sometimes in the form of offensive, triggering, or outrageous videos \cite{noauthor_youtube_nodate,haroon2023auditing}. 

In the context of both individual and societal reasons for users to better tailor their personalized content on YouTube, the platform provides several buttons such as ``Not interested'' and ``Don't recommend channel'', which allow users to express disinterest in a specific video or channel and alter their recommendation feeds accordingly \cite{burch_youtube_2019,cooper_how_2021}. These buttons exist among a variety of platform features, such as ``Disliking'' videos and deleting videos from one's watch history. All of those buttons may help users eliminate certain content from their feeds.

However, relatively little is known about the efficacy of these buttons in practice, nor about users' awareness of and confidence in them. To address these gaps, this work investigates how well simulated YouTube users (agents) can populate their recommendation feeds with content from a certain topic (the ``stain'' phase), as well as the ability for the recommendations of that topic to be removed by using a strategy to indicate disinterest (the ``scrub'' phase). We selected four topics: \emph{Alt-Right}, \emph{Antitheism}, \emph{Political Left}, and \emph{Political Right}. These topics are particularly interesting because, while plenty of literature exists on how one can get recommended more of these topics, little is known about removing them. Also, each topic is a realistic one that some users would no longer want to see. We examined six strategies (\emph{Watch neutral}, \emph{Dislike}, \emph{Delete}, \emph{Not interested}, \emph{No channel}, and \emph{Dislike recommended}), as well as a \emph{None} control strategy. Lastly, we conducted a complementary survey study to understand real users' awareness of and experience with those scrubbing strategies on YouTube.

The main findings are as follows:

\begin{itemize}[leftmargin=*]
    \item Watching a topic increased its presence on the homepage, though the stain never covers more than half of the recommendations. Watching a topic had less effect on the recommendations shown on videopages.
    \item For scrubbing a topic from the homepage, the most effective action was clicking the ``Not interested'' button on a recommended video. In contrast, none of the scrubbing significantly reduced the number of recommendations of videos on the topic shown on videopages.
    \item Nearly half of survey respondents were not aware of the most effective feature (pressing ``Not interested''). Those who were aware of it used it frequently, and perceived it to be less effective than the ``Don't recommend channel'' button, contrary to our findings from the audit study.
\end{itemize}
% ====================================================

% ====================================================
\section{Background and Related Work}
\label{sec:background}

\subsection{Locations of Relevant YouTube Features}

There are three main YouTube pages that are relevant to our study: the homepage, videopage, and watch history page. The \emph{homepage} is the landing page for users upon entering the platform.\footnote{\url{https://www.youtube.com}} It presents recommendations in a grid format. If a user is logged in, the content will be personalized to their account. Each recommendation contains a dropdown menu, where users are presented with two relevant buttons: the ``Not interested'' and ``Don't recommend channel'' buttons. These allow the users to (ostensibly) indicate disinterest with respect to specific recommendations.

The \emph{videopage} is the page users see while watching a video. Recommendations are given in a right-hand sidebar. The feature on this page that is relevant to our study is the ``Dislike'' button, indicated by a thumbs-down symbol.

Finally, the \emph{watch history page}\footnote{\url{https://www.youtube.com/feed/history}} (or \emph{watch history} for short) is the page that displays a log of the user's previously-watched videos. This page is only available to users who are logged in. Relevant to our study is the option for users to delete specific videos from their watch history. Specifically, the ``Delete'' button is ``X'' located on the upper-right hand corner of each video in the log. 

We note that these pages and features were accurately described at the time of data collection (August 2022) and the paper writing (January 2023). Since the platform often undergoes changes to its user experience and interface design, they may be outdated. 

\subsection{Sock Puppet Algorithm Audits}

Our study takes a sock puppet algorithm audit approach. Several prior works quantifying the YouTube recommender system's role in promoting and removing unwanted content also use the algorithm auditing approach. So we begin with a review of this method. 

An algorithm audit ``is a method of repeatedly and systematically querying an algorithm with inputs and observing the corresponding outputs in order to draw inferences about its opaque inner workings'' \cite{metaxa_auditing_2021}. Algorithm audits are a tool for researchers to investigate algorithms whose code and data are shielded from the public.

One type of algorithm audits is the sock puppet approach \cite{tomlein_audit_2021,haroon2023auditing,hosseinmardi2024causally}. Sock puppet audits use code scripts to create simulated users. These fake users -- also called ``agents'' -- interact with the platform or algorithm of interest as if they were real users. In the meantime, researchers record and compare the recommendations that the agents receive. 

\subsection{Recommender System's Role in Promoting Problematic and Unwanted Content}

YouTube is one of the most popular video sharing platforms. In recent years, it has received increasing public scrutiny from journalists and academics alike in assessing its recommendations of problematic content. 

The center of the platform's content dissemination is the recommendation engine, which plays an important role in helping users decide what to watch \cite{covington2016deep,wu2019estimating}. From a vast, growing pool of videos on the platform, users are suggested what to watch next based on their previous interactions with YouTube and what content they will most likely engage with next. \cite{zhao_recommending_2019}.

YouTube's recommendation engine has been theorized to promote problematic recommendations, which can broadly be split into two categories. The first is its propensity to suggest content that violates political, societal, and anti-democratic ideals such as extremism and conspiracy theories \cite{lewis_alternative_2018,tomlein_audit_2021} as well as political filter bubbles and radicalization \cite{tufekci_opinion_2018,hosseinmardi_examining_2021}. The second are those conflicting with individual preferences. Many users have found recommendations on video sharing platforms to be personally offensive, triggering, violent, and outrageous \cite{noauthor_youtube_nodate,haroon2023auditing}, as well as conflicting with their sense of identity \cite{karizat_algorithmic_2021}, even if the video is completely legal and enjoyable for others.

Several YouTube algorithm audits have investigated the role of recommender systems in promoting such content, largely focusing on political ideologies and conspiracy theories \cite{tomlein_audit_2021,haroon2023auditing,hosseinmardi2024causally}. While previous studies often refer to this phenomenon as ``filter bubbles'', we choose to use the term ``stain''. This is because previous studies (as well as ours) find that topical recommendations rarely take up more than half of one's feed and never reach 100\% despite watching many videos of that topic, and we would like to avoid the misleading interpretation of the term ``bubble'' as being completely surrounded by (i.e., having 100\% of) recommendations of a certain topic. 

Regardless of the term, studies have agreed that continued consumption of videos of a certain topic will lead to further (and sometimes increased) recommendation of that topic, on both the homepage and the videopage \cite{hussein_measuring_2020,papadamou2022just,haroon2023auditing}. Recommendations in search results, on the other hand, do not experience such personalization effects \cite{tomlein_audit_2021}. Despite these findings, it is still unknown if the stain is comprised completely of videos from previously-watched channels, or if YouTube introduces new channels of the topic yet to be watched by the user. Such information would demonstrate how much YouTube is recommending content beyond what is obviously related (i.e., that from the same channel), adding clarity to the current debate of the algorithm's role in information personalization.

Researchers have also studied other forms of problematic information personalization such as radicalization, or the process of being recommended content that is progressively more extreme \cite{ribeiro_auditing_2020,hosseinmardi_examining_2021,chen_exposure_2022}. By contrast, our study focuses on the construct of stain on YouTube. 

\subsection{User Controls to Remove Unwanted Content}

Combined calls from academics and journalists alike to mitigate the YouTube recommender's role in problematic content consumption have contributed to recent platform changes. These include features purportedly grant users more control in tailoring their recommendations \cite{burch_youtube_2019,cooper_how_2021}. Other social media platforms such as TikTok and Instagram have also released and experimented with similar user controls \cite{ariano_how_2021,noauthor_testing_2022}. Such features may improve user satisfaction in online spaces mediated by recommender systems, especially on systems (such as YouTube) that receive much of their content viewership from users watching recommender-suggested content.

However, compared to what is known about the prevalence of unwanted content on YouTube, relatively little is known about these features to remove them as well as their efficacy. We review this literature here.

\subsubsection{Algorithm audits on how to reduce recommendations.}
Two experiments used an intuitive strategy to try to reduce content of a given topic: watching videos of a \emph{different} topic. For example. \citet{tomlein_audit_2021}'s sock puppet audit found that agents were recommended less conspiratorial content after they watched many videos debunking conspiracy theories. \citet{haroon2023auditing}'s sock puppet audit found that a politically-biased recommendation feed could be ``debiased'' -- or achieve similar amounts of left and right-leaning videos -- by watching a diet of videos heavily featuring the ideology that was less prevalent originally. 

These studies suggest that it is possible to remove some unwanted content from one's feed. However, the degree to which it can be done varies and is never 100\%. 

Further, we find it necessary to expand recommendation-reduction strategies beyond video-watching and towards platform-provided buttons. First, many such buttons are designed for the explicit purpose of removing unwanted content (e.g., the ``Not Interested'' button). Second, they may be much faster to perform: studies suggest a minimum watch time of 10 minutes is required to register significant changes to recommendations \cite{papadamou2022just}; meanwhile, pressing a button takes just seconds. Lastly, using buttons may avoid the side effect of infusing too much content from another topic to replace the unwanted topic.

\citet{ricks_does_2022} provide the first quantitative study of such buttons on YouTube. They supplied users with a browser extension with a custom ``Stop Recommending'' button displayed on each video recommendation. Then, whenever it was clicked, users were randomly assigned to have their custom button press a native platform button in the background. Their results show that the native ``Don't recommend this channel'' button, which appears on recommendations, produced suggestions least similar to them.

\citet{ricks_does_2022}'s study benefits from a large sample. Their field experiment design also presents distinct advantages, particularly an external validity that a sock puppet audit cannot achieve. At the same time, we still find it valuable to perform a sock puppet experiment with a more controlled environment for two reasons. First, because users could press ``Stop Recommending'' on any recommendation from any topic, the study was not able to identify the effects of the buttons for well-defined topics. Second, there are possible confounds from uncontrolled user behavior, such as users watching videos similar to the ones they pressed ``Stop Recommending'' on, or cross-contamination between conditions where users clicked on YouTube-provided buttons in addition to Mozilla-provided ones.

\subsubsection{Users' relationship with user controls.}

A few studies also used qualitative methods to understand users' experiences and perceptions of different strategies to remove unwanted content from their personal feeds.

\citet{ricks_does_2022} surveyed and interviewed a subset of their participants from the quantitative arm of their study. They find that users take a variety of strategies to combat unwanted recommendations, generally find platform-provided features to be ineffective, and that achieving effective results require sustained time and effort.

While these surveys and interviews solicit the breadth of strategies that users have to combat unwanted recommendations, the degree to which general YouTube users are aware of each platform-provided feature is still unknown. It is also unknown whether they use these features, even if they are aware of them. Such data is important because an effective feature may be moot if they are unknown or unused. 

\citet{smith_dark_2021} also examined YouTube user controls for altering recommendations. They found that the actions performed by such buttons were reactive (i.e., only useful \emph{after} a user received an unwanted recommendation) and that the feedback provided to the user after clicking them was often unclear and vague. They also found that navigating to some of these features was difficult, which could limit users' ability to use them.
% ====================================================

% ====================================================
\section{Research Questions}
\label{sec:research_questions}

Previous studies of ``filter bubbles'' on YouTube see that recommendations of a given topic can increase as a result of watching videos of that topic, but we do not know whether these recommendations are from channels the user has watched before, or whether they are new channels that YouTube finds similar. Such a breakdown would add to the knowledge of YouTube's role in promoting unwanted content by quantifying how much YouTube is ``inferring'' this content rather than simply suggesting content from previously-watched channels. Also, confirming the general result of increasing topical recommendations would motivate our next study phase, which attempts to remove them. 

Thus, we first address the question, \textbf{how responsive are YouTube recommendations to watching many videos of the same topic?} (RQ1) In particular, do they recommend more videos of the same topic, and if so are they from channels that users watched up to that point or are they new ones? Are they different for different topics? We study four topics whose prevalence on YouTube has been previously studied: \emph{Alt-Right}, \emph{Antitheism}, \emph{Political Left}, and \emph{Political Right} (motivated and described in \Cref{ssec:topics}). 

We are also interested in the effects of platform features in removing unwanted recommendations. While a previous study investigated their usage ``in the wild'', the effects of each feature, uncontaminated by usage of other features, on topics that are well-defined, is still unknown. Such questions are worth answering because YouTube users in general could benefit from knowing what the most effective strategies for removing unwanted content are, specifically their effect on specific topics that they may dislike. 

Thus, we ask, \textbf{how responsive are YouTube recommendations to repeatedly performing a particular strategy to try and remove unwanted videos of a topic?} (RQ2) Are they different between videopage and homepage? How much content is removed from similar channels that are not explicitly interacted with? Do they vary topic to topic? We identified six such strategies, such as pressing the ``Not interested'' button, and listed them in \Cref{ssec:strategies}. 

Finally, it is unknown how many YouTube users are aware of each platform feature, how many utilize them, and how effective the users find them to be. This information is important because effective strategies may be moot if users do not know their existence, and because users should be both using effective strategies and finding them to be effective.

Therefore, we lastly ask, \textbf{what are real users' experiences with the platform features that we test in RQ2?} (RQ3) With respect to each platform feature, we designed a survey study to ask how many participants are aware of it, what percentage use it to remove unwanted recommendations (given they are aware), and how effective participants find it to be (given they are aware and have used the feature to try to amend the situation).
% ====================================================

% ====================================================
\section{Sock Puppet Study}
\label{sec:methods}

\subsection{Sock Puppet Design}

We take a sock puppet algorithm audit approach to examine how suggestions from certain topics can both be populated onto and removed from one's personal recommendation feed. Broadly, our agents first purposely populate their feed with videos from this unwanted topic (``stain phase''); Then, they take on one of a variety of strategies to try to eliminate such videos from being recommended (``scrub phase''). We collect data on how recommendations change throughout these phases in order to characterize the recommendation system's response to these various interactions. 

\subsubsection{Video topics.} 
\label{ssec:topics}

We require video topics as an input for our agents to populate in their recommendations (staining phase), and then attempt to scrub (scrubbing phase). 

Each topic is operationalized as a list of channels collected by researchers who have formerly studied that topic on YouTube. They are used in our experiment in two ways. First, agents watch videos from the channel lists during the stain phase. Next, during the scrub phase, some strategies cross reference their homepage recommendations with the assigned topic's channel list to determine whether and which one to indicate disinterest on. 

\begin{itemize}
    \item \emph{Alt-Right}: The most extreme group of the Alternative Influence Network, a loosely-defined community of YouTube channels that are defined by their opposition to mainstream media \cite{ricks_does_2022}. The Alt-Right promotes white nationalism in the face of an increasingly diverse US population, and is often openly anti-semitic \cite{noauthor_alt_2019}. YouTube channels of the Alt-Right were first collected by \citet{lewis_alternative_2018} through a snowballing method, and subsequently augmented by \citet{ribeiro_auditing_2020} and \citet{chen_exposure_2022}. 
    \item \emph{Antitheism}: Collected by \citet{ledwich_algorithmic_2019}. It is ``the self-identified atheist who is also actively critical of religion''. 
    \item \emph{Political Left}: Collected by \citet{wu_cross-partisan_2021}. They include local news, talk shows, and magazines. We use the US political left channels, which takes similar views among various issues of political significance, such as climate change.
    \item \emph{Political Right}: Same as above, but with the US political right channels.
\end{itemize}

\subsubsection{Scrubbing strategies.}
\label{ssec:strategies}

The name and operation of each scrubbing strategy are listed below. Each agent is assigned one strategy, and performs it repeatedly during the ``scrub phase'' of the sock puppet run.

\begin{itemize}
    \item \emph{None} (control): Load the homepage, then do nothing except refresh the homepage.
    \item \emph{Watch neutral}: Load and watch a video from mainstream, politically neutral news outlets as defined by the fact-checking organization Media Bias/Fact Check.\footnote{\url{https://mediabiasfactcheck.com/}}
    \item ``History-based'' strategies
    \begin{itemize}
        \item \emph{Dislike}: Load a previously-watched video from the stain phase and click the ``Dislike'' button.
        \item \emph{Delete}: Load the watch history and click ``Delete'' on the most recently-watched video. 
    \end{itemize}
    \item ``Recommendation-based'' strategies. Load the homepage. If there does \emph{not} exist any recommended video on the homepage from a channel in the channel list, then just refresh. However, if such a video exists, do the following to the first such video:
    \begin{itemize}
        \item \emph{Not interested}: click the ``Not interested'' button and refresh the homepage.
        \item \emph{No channel}: click the ``Don't recommend channel'' button and refresh the homepage.
        \item \emph{Dislike recommended}: click on the video and dislike it (agents do not stay to watch the video), then return to the homepage.
    \end{itemize}
\end{itemize}

The ``watch neutral'' strategy attempts to ignore the current issue by watching videos from a different topic, and most resembles the intervention strategies of related studies \cite{haroon2023auditing,tomlein_audit_2021}. We call dislike and delete strategies ``history-based'' because they act on videos that the agents watched during the stain phase. We call the final three strategies ``recommendation-based'' because they are performed with respect to recommended videos.

\subsubsection{Sock puppet phases and data collection.}

A sock puppet agent follows the following process. After logging in to a YouTube account, an agent performs the ``stain phase'', where it watches 40 videos, for up to 30 minutes each,\footnote{An alternative is to stay for the median watch time for videos with similar length, see \citet{wu2018beyond}'s computation of relative engagement metric.} from a ``stain video list'' which are sampled from the channel list belonging to its assigned topic. Next, it performs the ``scrub phase'', where it executes its assigned scrubbing strategy 40 times. Lastly, the agent clears its entire YouTube activity through Google's MyActivity page,\footnote{\url{https://myactivity.google.com/myactivity}} in order to leave a clean history for the next audit to start \cite{tomlein_audit_2021}. This includes clearing all revertible actions made during its run, such as clicks of the “Dislike”, “Not interested”, “Don’t recommend”, and “Delete from watch history” buttons. 

Our agents use web-scraping methods to collect the top 10 recommendations from the homepage and videopage at three strategic points: 
\begin{itemize}
    \item P1: The beginning of the stain phase.
    \item P2: The end of the stain phase.
    \item P3: The end of the scrub phase.
\end{itemize}

Because the video being watched during videopage collection may itself have an effect on recommendations, each agent always loads the \textit{same} video at all three collection points (P1, P2, P3). This video is from the stain video list.

\begin{figure*}[tbp]
   \centering
   \includegraphics[width=1\linewidth]{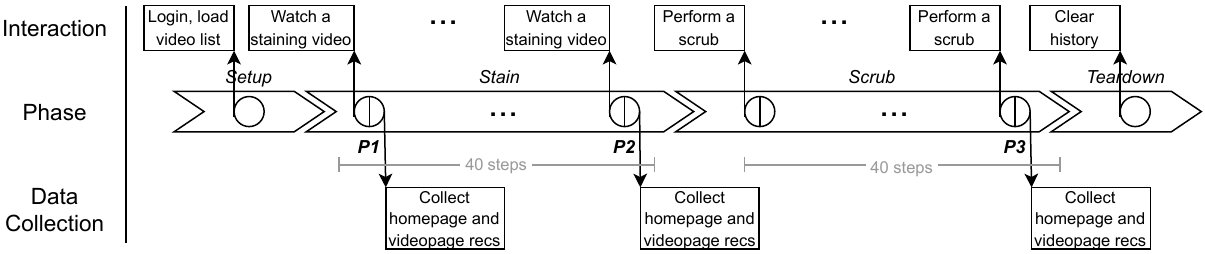}
   \caption{The interactions and data collection that a sock puppet performs with the YouTube platform, broken down by phase.}
   \label{fig:sock_puppet_process}
\end{figure*}

Altogether, \Cref{alg:sock_puppet} provides an overview of a sock puppet agent's interactions with the platform. A visual flowchart of this process is given in \Cref{fig:sock_puppet_process}.

% final submission: include flowchart

% ----------------------------------------------------
\begin{algorithm}
    \caption{Agent}%Agent process (``rec'' = ``recommendation'')}
    \begin{algorithmic}
        \State Log into YouTube
        \State Collect homepage recs \Comment{P1}
        \State Collect videopage recs from the first stain video \Comment{P1}
        \For{$i \in [2 \dots 40]$} \Comment{stain phase}
            \State Watch a video from stain video list up to 30 minutes
        \EndFor
        \State Collect homepage recs \Comment{P2}
        \State Collect videopage recs from the first stain video \Comment{P2}
        \For{$i \in [1 \dots 40]$} \Comment{scrub phase}
            \State Perform assigned scrubbing strategy
        \EndFor
        \State Collect homepage recs \Comment{P3}
        \State Collect videopage recs from the first stain video \Comment{P3}
        \State \sw{Clear YouTube activity (cancel all revertible actions)}
        \label{alg:sock_puppet}
    \end{algorithmic}
\end{algorithm}
% ----------------------------------------------------

% ----------------------------------------------------
% ----------------------------------------------------

We now describe the configurations of agents for the overall experiment. For each topic we tested seven strategies, each five times, resulting in (7 * 5 =) 35 agents. All 35 agents of a given topic were run in parallel in order to deal with recommendation noise that may arise from having agents make queries at different times.

For a given topic, we also drew five stain video lists, and assigned each list to exactly one agent within every strategy tested. Doing so assures that agents of the same strategy watch different sets of staining videos, boosting generalizability of the strategy effects, while simultaneously assuring that agents of different strategies watch, in total, the same videos, enabling comparability between strategies. 

Additionally, each of the 35 agents have their own Google Accounts so that the platform can track their viewing habits and personalize content to each agent, and so that we can more closely simulate real users' experience with the platform. Logging into an account also grants access to buttons and features that are only available to users that are logged in (e.g., the ``Not interested'' button).

Our agents ran in a Google Chrome browser with adblocker installed. They had Google accounts with birthdays set at an arbitrary 5/5/1990, a gender selection of ``Rather not say'', and asexual names (e.g., ``Tandy''). We also address the potential biases from location effects, which would occur if queries were made to the platform from different locations, or from (different accounts in) the same IP address. Thus, all agents are created and live in the same AWS Region of Ohio (US-East-2), but make queries from individual IP addresses.

Out of 140 sock puppets released over the course of five days in August 2022, 139 sock puppets ran successfully. Agents collected a total of 8330 recommendations.

\subsection{Data Annotation}

Our agents collected many recommendations during their runs. We would like to label them for whether they belong to the topics that the agents were assigned to (what we call ``stain''), in order to quantify how well (1) the stain phase worked to populate agent' recommendations with stain, and (2) how well the scrub phase worked to remove it.

We adopted an iterative strategy in annotating the recommended channels. We first developed an initial annotation codebook by surveying prior research. Next, we randomly sampled 50 channels for each topic. Two authors who had extensive experience in studying political polarization and YouTube platforms independently labeled those channels by following the codebook. The preliminary inter-rater reliability (IRR), measured by Cohen's kappa, was 0.648, 0.728, 0.634, 0.563 for \emph{Alt-Right}, \emph{Antitheism}, \emph{Political Left}, \emph{Political Right}, respectively, demonstrating substantial agreement. The two raters discussed every disagreed case to reach consensus and updated the codebook whenever needed. 

The two raters then went on and labeled all the remaining channels. The final IRR kappa scores were 0.660, 0.822, 0.854, and 0.945, respectively. The raters also discussed all disagreed cases and resolved disagreement. The final annotation codebook is attached in~\Cref{app:codebook}. The annotation results and IRR calculation can be previewed via this link.\footnote{Anotation results: \url{https://tinyurl.com/45t2rhrs}}

% ----------------------------------------------------
\setlength{\tabcolsep}{3pt}
\begin{table*}[ht]
\small
\centering
\begin{tabular}{
    r | r |
    ccccc | 
    ccccc |
    ccccc |
    ccccc |
    cc
}
     &  & \multicolumn{5}{c}{\emph{Alt-Right}} & \multicolumn{5}{c}{\emph{Antitheist}} & \multicolumn{5}{c}{\emph{Political Left}} & \multicolumn{5}{c}{\emph{Political Right}} & Avg. relative change \\
    & & P1   &    & P2    &    & P3   & P1    &    & P2    &    & P3   & P1   &   & P2   &   & P3   & P1   &    & P2   &   & P3   & P2 to P3       \\ \hline
\multirow{8}{*}{\rotatebox[origin=c]{90}{Homepage}} & Total          & 2\%  & *  & 20\%  &    &      & 11\%  & *  & 37\%  &    &      & 6\%  & * & 28\% &   &      & 9\%  & *  & 29\% &   &      &     \\ 
& \emph{None}    &      &    & 20\%  &    & 22\% &       &    & 32\%  &    & 38\% &      &   & 20\% &   & 34\% &      &    & 32\% &   & 42\% & +32\%      \\ 
& \emph{Watch neutral}          &      &    & 12\%  &    & 8\%  &       &    & 34\%  &    & 22\% &      &   & 32\% & * & 18\% &      &    & 26\% & * & 16\% & -38\%     \\ 
& \emph{Delete}         &      &    & 20\%  & *  & 0\%  &       &    & 32\%  &    & 24\% &      &   & 30\% & * & 0\%  &      &    & 24\% & * & 4\%  & -77\%     \\ 
& \emph{Dislike}        &      &    & 20\%  &    & 4\%  &       &    & 30\%  & *  & 10\% &      &   & 24\% &   & 28\% &      &    & 32\% & * & 16\% & -45\%     \\ 
& \emph{Not interested} &      &    & 18\%  & *  & 0\%  &       &    & 42\%  & *  & 2\%  &      &   & 28\% & * & 10\%  &      &    & 26\% & * & 2\%  & -88\%     \\ 
& \emph{No channel}     &      &    & 28\%  & *  & 10\%  &       &    & 44\%  & *  & 22\% &      &   & 30\% & * & 16\%  &      &    & 36\% & * & 24\% & -49\%     \\ 
& \emph{Dislike rec.}   &      &    & 20\%  & *  & 8\%  &       &    & 42\%  & *  & 14\%  &      &   & 32\% &   & 24\% &      &    & 26\% &   & 34\% & -30\%      \\ \hline
\multirow{8}{*}{\rotatebox[origin=c]{90}{Videopage}} & Total    & 6\%     &    & 5\%     &    &    & 26\%     & *    & 37\%     &    &    & 41\%        &       & 42\%        &       &       & 12\%    & *      & 28\%    &        &        &          \\ 
&\emph{None}    &    &    & 6\%     &    & 4\%     &    &    & 40\%     &    & 34\%     &       &       & 38\%        &       & 34\%        &        &        & 20\%    &        & 30\%    & -2\%     \\ 
&\emph{Watch neutral}          &    &    & 4\%     &    & 6\%     &    &    & 28\%      &    & 6\%      &       &       & 42\%        &       & 46\%        &        &        & 30\%    &        & 34\%    & -1\%      \\ 
&\emph{Delete}         &    &    & 2\%     &    & 2\%     &    &    & 38\%     &    & 34\%     &       &       & 42\%        &       & 34\%        &        &        & 32\%    &        & 16\%    & -20\%      \\ 
&\emph{Dislike}        &    &    & 12\%      &    & 14\%      &    &    & 40\%     &    & 34\%      &       &       & 44\%        &       & 46\%        &        &        & 22\%    &        & 42\%    & +24\%     \\ 
&\emph{Not interested}         &    &    & 4\%     &    & 2\%     &    &    & 40\%     &    & 48\%     &       &       & 44\%        &       & 44\%        &        &        & 32\%    &        & 30\%    & -9\%      \\ 
&\emph{No channel}     &    &    & 0\%     &    & 0\%     &    &    & 32\%     &    & 30\%     &       &       & 44\%        &       & 40\%        &        &        & 34\%    &        & 36\%    & -2\%     \\ 
&\emph{Dislike rec.} &    &    & 6\%     &    & 2\%     &    &    & 42\%     &    & 34\%     &       &       & 40\%        &       & 50\%        &        &        & 24\%    &        & 22\%    & -17\% 
\end{tabular}

\caption{Stain percentage values, and Wilcoxon signed-rank tests, for the homepage (top) and videopage (bottom), at data collection points P1, P2, P3, for each strategy (row) and topic (column). Values in the ``Total" row combine stain for all agents of that topic. The star (*) between two values represents significant changes at the 0.05 significance level.}
\label{table:signed_test}
%}
\end{table*}
% ----------------------------------------------------

% ----------------------------------------------------
\begin{table*}[ht]
\centering
\small
\begin{tabular}{r|m{1cm}m{1cm}m{1cm}|m{1cm}m{1cm}m{1cm}|m{1cm}m{1cm}m{1cm}|m{1cm}m{1cm}m{1cm}}
          & \multicolumn{3}{c}{\emph{Alt-Right}}              & \multicolumn{3}{c}{\emph{Antitheism}}             & \multicolumn{3}{c}{\emph{Political Left}}          & \multicolumn{3}{c}{\emph{Political Right}}         \\
          & Off-topic & On-topic watched & On-topic new & Off-topic & On-topic watched & On-topic new & Off-topic    & On-topic watched & On-topic new & Off-topic     & On-topic watched & On-topic new \\ \hline
Homepage  & 79\%       & 16\%   & 5\%                  & 60\%       & 18\%   & 23\%                 & 72\%          & 2\%    & 26\%                 & 70\%           & 9\%    & 21\%                 \\ 
Videopage & 95\%       & 5\%    & 0\%                  & 63\%       & 9\%    & 29\%                 & 58\%          & 13\%   & 29\%                 & 72\%           & 19\%   & 9\%                 
\end{tabular}

\caption{Stain at P2 split into categories of off-topic, on-topic watched-channel, and on-topic new-channel (we omit the word ``channel" to save space), for each of homepage and videopage (row) and each topic (column).}
\label{table:p2}
\end{table*}
% ----------------------------------------------------

\subsection{Result 1: Stain Phase} 
\label{results_stain}

In this subsection, we answer our questions posed in RQ1.

\subsubsection{Effects of stain phase.}

We wanted to know whether our agents experienced a significant change in stain (the percentage of recommendations of their assigned topic) after the stain phase. To address this question, we compared the stain of our agents at P1 with those at P2 for each topic, in both the homepage and the videopage. 
To determine whether changes were significant, we chose the Wilcoxon signed-rank test because (a) the data was non-normal; (b) the comparison before and after the ``stain phase'' treatment was a paired test.
Results are given in Table \ref{table:signed_test} (P1 to P2).

In the homepage, we find that all topics experienced significant increases in stain as a result of the stain phase. \emph{Antitheism} received the most stain at P2 (37\%), while \emph{Alt-Right} received the least (20\%). In contrast, the videopage demonstrated significant changes in stain only on \emph{Antitheist} and \emph{Political Right}. \emph{Alt-Right} actually showed a slight decrease from P1 to P2. Despite this, a non-zero stain still existed in the videopage -- absolute percentages at P1 varied between 5\% for \emph{Alt-Right} and 42\% for \emph{Political Left}. We lastly remark that, across both homepage and videopage, and across topics and strategies, stain never reached more than half. 

These findings set us up well for the scrub phase, because it assures that our agents will indeed have stain to remove when they perform their scrubbing strategies.

% ----------------------------------------------------
\begin{table*}[t]
\centering
\small
\begin{tabular}
{r|m{1cm}m{1cm}m{1cm}|m{1cm}m{1cm}m{1cm}|m{1cm}m{1cm}m{1cm}|m{1cm}m{1cm}m{1cm}}
          & \multicolumn{3}{c}{\emph{Alt-Right}}                                & \multicolumn{3}{c}{\emph{Antitheism}}                               & \multicolumn{3}{c}{\emph{Political Left}}                            & \multicolumn{3}{c}{\emph{Political Right}}                           \\
               & Off-topic & On-topic scrubbed & On-topic new & Off-topic & On-topic scrubbed & On-topic new & Off-topic & On-topic scrubbed & On-topic new & Off-topic & On-topic scrubbed & On-topic new \\ \hline
\emph{Watch neutral}          & 90\%      & 8\%               & 2\%          & 74\%      & 14\%              & 12\%          & 78\%      & 2\%               & 20\%         & 82\%      & 6\%               & 12\%         \\ 
\emph{Delete}         & 96\%     & 0\%               & 4\%          & 74\%      & 18\%              & 8\%         & 100\%     & 0\%               & 0\%          & 94\%      & 2\%               & 4\%          \\ 
\emph{Dislike}        & 92\%      & 2\%               & 6\%          & 84\%      & 10\%              & 6\%          & 72\%      & 0\%               & 28\%         & 84\%      & 2\%               & 14\%         \\ 
\emph{Not interested} & 100\%     & 0\%               & 0\%          & 96\%      & 0\%               & 4\%          & 90\%      & 0\%               & 10\%          & 94\%     & 0\%               & 6\%          \\ 
\emph{No channel}     & 84\%      & 0\%               & 16\%          & 76\%      & 2\%               & 22\%         & 84\%      & 0\%               & 16\%          & 76\%      & 0\%               & 24\%         \\ 
\emph{Dislike rec}.   & 90\%      & 2\%               & 8\%          & 80\%      & 0\%               & 20\%          & 76\%      & 0\%               & 24\%         & 64\%      & 4\%               & 32\%        
\end{tabular}

\caption{Stain at P3 on the homepage split into categories of off-topic, on-topic scrubbed-channel, and on-topic new-channel (we omit the word ``channel" to save space), for each of strategy (row) and each topic (column).}
\label{table:p3}
\end{table*}
% ----------------------------------------------------

\subsubsection{Stain from watched channels vs. new channels.}

We wanted to examine the stain in P2 with more granularity. Specifically, how much of it was from channels the agent had explicitly watched before, and how much was from channels that the agents had never explicitly watched before? 

To answer this question, we categorized all recommendations at point P2 as ``off-topic'', ``on-topic watched-channel'' (i.e., the agent that collected this recommendation had already watched a video from the same channel during the stain phase), or ``on-topic new-channel'' (i.e., the agent had not watched any videos from the same channel up to that point). Then, for each topic, we found the percentage of recommendations belonging to each category. We report these ratios for the homepage and videopage in Table \ref{table:p2}.

On the homepage, we find that \emph{Political Left} had the most recommendations from new channels (26\%), as well as the highest ratio of recommendations from new channels to those from watched channels (13:1).
On the other hand, the \emph{Alt-Right} had the smallest new-channel percentage, both in absolute terms (5\%) and as a ratio of watched-channel percentage ($\sim$1:3).

On the videopage, \emph{Political Left} and \emph{Antitheist} had the highest absolute percentage of recommendations from new channels (29\%), and \emph{Antitheist} had the highest ratio of new channel recommendations to those from watched channels ($\sim$3:1). By contrast, the \emph{Alt-Right} agents received no recommendations from new channels (0\%) while all other topics received at least 9\%.

These findings suggest that the YouTube recommendation system sometimes plays a role in providing stain to the user by not only suggesting content that is from the same channel, but rather by inferring and providing that from different but similar channels.

\subsection{Result 2: Scrub Phase} 
\label{results_scrub}

In this subsection, we answer our questions posed in RQ2.

\subsubsection{Effects of scrub phase.}

We wanted to know whether our agents could remove stain after the scrub phase. To address this question, we compared the stain of our agents at P2 with those at P3, for each topic, in both homepage and videopage. Again, our data was non-normal and paired, so we ran Wilcoxon signed-rank tests to see whether stain decreased significantly. Results are again in~\Cref{table:signed_test} (P2 to P3).

On the homepage, \emph{Not interested} and \emph{No channel} were the only strategies that significantly reduced the amount of stain across all topics. Comparing average relative changes between P2 and P3, \emph{Not interested} wins out (-88\%). One strategy successfully scrubbed three out of four topics (\emph{Delete}), while two strategies were successful in two out of four topics (\emph{Dislike recommendation} and \emph{Watch neutral}). On the other end, the \emph{None} strategy did not produce any significant effect (in fact, it produced a slight increase), which was expected because it was our control strategy. 
On the videopage, we did not find any significant scrub phase effects.

\subsubsection{Stain from scrubbed channels vs. new channels.}

We wanted to know whether scrubbing strategies removed stain in general, or if they only removed the subset of channels the agent had explicitly scrubbed up to that point.

To answer this question, for all recommendations at point P3, we categorized them as either ``off-topic'', ``on-topic scrubbed-channel'' (i.e., the agent that collected this recommendation had already scrubbed a video from the same channel during the scrub phase), or ``on-topic new-channel'' (i.e., the agent had not scrubbed a video from the same channel up to that point). Notice that these categories are analogous to watched/new categories made for P2 in Section \ref{results_stain}. 

Then, for each topic/strategy pairing, we found the ratio of recommendations at P3 that belonged to each category. We report these ratios for the homepage in Table \ref{table:p3}. We did not examine the videopage because it did not experience any significant changes in this phase. The \emph{None} strategy was excluded because no videos were scrubbed. 

Categorizing recommendations this way reveals that scrubbing strategies behaved differently in removing unwanted recommendations. For instance, in three out of four topics, at least half of the \emph{Watch neutral} strategy's remaining stain at P3 was from scrubbed channels. On the other hand, \emph{No channel} rarely left recommendations from scrubbed channels (0-2\%); most stain remaining after using this strategy was from new channels. The behavior of \emph{No channel} agrees with many user perceptions of the ``Don't recommend channel'' button \cite{ricks_does_2022}, and matches an intuitive interpretation of the button name. 
% ====================================================

% ====================================================
\section{Survey Study}
\label{sec:survey}

\subsection{Survey Design}

In the sock puppet section of our work, we collected data from simulated users' interactions with YouTube to quantify how platform features may help remove unwanted recommendations. In this section, we want to understand better the relationship between real users and these features. We ran a survey to determine this.

% ----------------------------------------------------
\begin{table*}[ht]
\small
\centering
\begin{tabular}{r|ccc}
  & \emph{Awareness}             & \emph{Usage}                  & \emph{Belief in efficacy} \\ 
  \hline
  Delete   & 51.41\% $\pm$ 7.55\% {[}248{]} & 53.64\% $\pm$ 13.52\% {[}110{]} & 3.76 $\pm$ 0.25 {[}48{]}  \\
  Dislike  & 93.94\% $\pm$ 3.90\% {[}263{]} & 37.75\% $\pm$ 7.90\% {[}226{]}  & 2.52 $\pm$ 0.25 {[}62{]}  \\
  Not interested & 56.03\% $\pm$ 6.63\% {[}258{]} & 82.83\% $\pm$ 8.23\% {[}156{]}  & 3.42 $\pm$ 0.32 {[}122{]} \\
  Don't recommend channel & 35.37\% $\pm$ 5.85\% {[}255{]} & 80.53\% $\pm$ 9.58\% {[}111{]}  & 4.10 $\pm$ 0.33 {[}88{]}  
\end{tabular}
\caption{Results from user survey. For each button (row), we list the point estimate and confidence level of constructs of interest (column). Sample sizes are given in brackets. Note that \emph{Awareness} and \emph{Usage} are in the range of 0 to 100\% while \emph{Belief in efficacy} is in a scale of 1 to 5.}
\label{table:survey}
\end{table*}
% ----------------------------------------------------

\subsubsection{Survey overview.}

We first asked whether respondents had experienced getting unwanted recommendations before. Respondents were specifically asked whether they have experienced this scenario before: ``\emph{You are browsing YouTube, and notice videos recommended to you that you would rather not have recommended (because they are offensive to you, triggering, not safe for work, or some other reason)}''. The buttons we consider are ``Delete'' (delete a video from watch history), ``Dislike'', ``Not interested'', and ``Don't recommend channel''. We asked for respondents' experiences with these buttons, with respect to three constructs: 

\begin{itemize}
    \item \emph{Awareness}: Before taking this survey, were you aware this button existed?
    \item \emph{Usage}: Have you used this button to remove unwanted recommendations?
    \item \emph{Belief in efficacy}: Recall the times when you used this button to remove unwanted recommendations. How effective do you think it was? Please rate from 1 (not at all effective) to 5 (completely effective).
\end{itemize}

Only those who have experienced unwanted recommendations before and were aware of the buttons were asked to report on their real usage, while others were given a hypothetical question (``\textit{If} you had known this button existed, would you have used it?"). Furthermore, only those who were (1) experienced, (2) aware of button, and (3) have used the button were asked to report their belief in its efficacy in removing recommendations, while others were given a hypothetical question (``\textit{If} you had used this button for this scenario, how effective do you think would you find it?").

\subsubsection{Survey implementation.}
We recruited 300 participants from the survey recruitment platform Prolific, and ran the survey on the survey delivery platform Qualtrics. We selected participants that had used YouTube before, were adults (18+), and resided in the US. Participants were paid \$15 an hour. The University of Michigan Health Sciences and Behavioral Sciences Institutional Review Board has determined that this research is exempt from IRB oversight (Study ID: HUM00224551).

Surveys were pretested with colleagues. We emphasized honest rather than ``right'' answers so that respondents would not be tempted to ``please'' us by saying they knew about a button when in reality they did not \cite{paolacci_inside_2014}. We included screenshots of buttons so that they didn't have to know them by name. Since attention checks are important to maintaining experimental validity, we also implemented three of them throughout the survey to make sure the respondents were focusing on and comprehending the survey questions. At three separate points, we gave them a question whose format was identical to that of others and instructed them to select a specific choice. For example, \emph{``Please select `Dislike' in the choices below"}.

We filtered responses by eliminating one from those who failed any of the three attention checks. Respondents were paid regardless of whether they passed or failed attention checks. Then, we eliminated responses from anybody who answered ``not sure" to our questions.

\subsection{Survey Analysis Methods and Results}
In this subsection, we answer our questions posed in RQ3.

In total, our survey received 274 responses from those who passed all three attention checks. However, our respondent sample did not immediately generalize to a more general population. Thus we used post-stratification, a popular statistical method that adjusts estimates on non-probability samples \cite{salganik_bit_2019}, to generalize our results to the adult YouTube-using population in the US. 

To perform post-stratification, we divided our respondents into binary genders and age buckets (roughly 20 years apart), making a total of eight subgroups. We then made estimates of each subgroup's prevalence in the target population by combining Census data on age/gender subgroups \cite{duffin_population_2022} and PEW data on the percentage of each subgroup that use YouTube \cite{auxier_social_2021}. Comparing our survey sample's distribution among subgroups with that of the target population revealed that our sample skewed young: Among usable responses, we routinely over-sampled the 18-45 subgroups and under-sampled 65+ ones. Fortunately, post-stratification corrects this bias.

We report the answers in Table \ref{table:survey}. \emph{Awareness} percentages were calculated by aggregating answers from all respondents that passed our attention checks. For \emph{Usage}, we restricted our calculation to those who were both aware that the feature existed, and had experienced having unwanted recommendations. \emph{Belief in efficacy} was the most restricted because we only wanted ratings from those who would be well-informed of its effects from personal usage: Only those who experienced unwanted recommendations, were aware the button existed, and used that button to try to resolve the issue, were considered. These population restrictions are applied to both the table and our discussion of results.

Moving onto results, we find for \emph{Awareness} that survey respondents were most aware of the ``Dislike" button's existence (93.94\%). ``Don't recommend channel" was the least well-known (35.37\%). As for \emph{Usage}, they favored ``Not interested" (82.83\%) and ``No channel" (80.53\%) to remove unwanted recommendations when they experienced it. ``Dislike" was the least used button (37.75\%). Looking at \emph{Belief in efficacy}, users found ``Delete" (3.76), ``Not interested" (3.42), and ``No channel" (4.10) all more effective than the ``Dislike" button (2.52). 

These findings suggest that users do not use the ``Dislike" button to remove unwanted recommendations, despite most knowing about its existence. Respondents' intuition of this button match our empirical findings: We saw in \Cref{results_stain} that the \emph{Dislike} and \emph{Dislike recommendation} strategies both reduced less stain compared to other scrubbing strategies (\emph{Delete}, \emph{Not interested}, \emph{No channel}). Meanwhile, the button for our (empirically) most effective scrubbing strategy -- ``Not interested" -- was highly used by respondents who knew of it. However, awareness was a privilege: almost 44\% of survey respondents were unaware of its existence.
% ====================================================

% ====================================================
\section{Discussion}
\label{sec:discussion}

We performed an algorithm audit of the YouTube recommendation system to test whether one could remove unwanted content from their feed. We paired our audit with a survey to understand whether users actually knew these buttons existed, used them, and believed them to be effective.

In our audit, we found that the stain phase produces a significant increase in stain in the homepage across all topics. We also saw that stain at P2 never reached more than half of recommendations in either the homepage or the videopage. These results confirm our suspicion that watching many videos from a given topic do not completely ``surround" the user topical recommendations like the term ``filter bubble" would suggest, and motivated our usage of the term ``stain".

Continuing on results from the stain phase, we broke down stain at P2 into those from channels watched before and those from channels not watched before, finding that their prevalence varied based on topic. 

We saw that both types of stain were present, but to varying degrees depending on topic. On the one hand, \emph{Political Left} received the most stain from new channels for both the homepage and videopage, demonstrating that the platform had a notion of topical similarity by ``inferring" other channels from the political left that the user may like. On the other hand, the \emph{Alt-Right} received the least recommendations from new channels, for both the homepage and videopage. This finding is interesting given YouTube's recent public promises to curb misinformation and conspiracy theories \cite{noauthor_continuing_2019}, especially "harmful" ones such as Q-Anon \cite{noauthor_managing_2020}, as well as a shift in company-wide attention towards stopping home-grown, right-wing extremism from spreading on its platform \cite{bergen_youtube_2022}. While the lack of recommendations from new \emph{Alt-Right} channels supplied to agents who watch that content could be evidence of YouTube operationalizing its promises, we cannot formally tell the difference between that and a general lack of \emph{Alt-Right} videos remaining on the platform today.

Moving onto the scrub phase, we compared different scrubbing strategies and found that \emph{Not interested} was the most effective one on the homepage: It produced significant decrease in stain across all topics, and using it resulted in the greatest average decrease in stain from P2 to P3 across topics (-88\%). This strategy performed well in removing stain from both channels it had explicitly scrubbed as well as similar ones it didn't interact with. Thus, users who would like to remove recommendations from any channels belonging to an unwanted topic should use this strategy.

In contrast to homepage findings, we found that the videopage never experienced significant effects from the scrub phase. At a cursory glance, it seems that our results disagree \citet{tomlein_audit_2021}'s finding: in their study, agents \textit{could} significantly reduce conspiratorial recommendations on the videopage by watching many videos debunking the conspiracy theory. However, upon further inspection it should be noted that in fact we have two separate experiments. Whereas agents in our study collected videopage recommendations from a video at P3 that was the same as that used in P2, their study used a video at P3 that was the semantic \emph{opposite} of that of P2. Specifically, \citet{tomlein_audit_2021}'s bots collected them from a video \emph{promoting} agents' assigned conspiracy theory at P2, and then collected them from a video \emph{debunking} it at P3.

Combining our findings with those from \citet{tomlein_audit_2021} suggests that videopage recommendations may be influenced more by the video that is playing while they are collected, than any interactions with the system leading up to that collection. The implication for users is that they should not expect any scrubbing strategies to save them from further recommendations of an unwanted topic if they plan to keep watching a video of that topic; Rather, they may want to stop watching content from that topic altogether. 

Lastly, we wanted to know how users interacted with platform features in their daily YouTube usage. We found that US adult YouTube users were most aware of the ``Dislike" button, yet more empirically effective strategies, such as ``Not interested", were lesser known. Those who knew the ``Not interested" button existed used it at a higher rate and saw it as more effective than those who knew of ``Dislike".

Put together, our sock puppet and survey findings suggest that if YouTube wanted to allow users to more effectively remove unwanted recommendations, it should make its effective platform-features for doing so more broadly known to the general YouTube population. Doing so would not only benefit users' experience; it would also be in the best interest of the platform because allowing users to have more agency to tailor algorithmic decisions to their preferences can build and maintain their trust in the system \cite{ekstrand_letting_2015}, as well as increase overall satisfaction \cite{shin_how_2020}. One implication for platform designers is that they should make buttons such as ``Not interested" more widely known by increasing its discoverability on the website. To that end, \citet{ricks_does_2022} provide a blueprint. In their experiment, they found that when their users were displayed ``Don't recommend this" buttons prominently and clearly on recommendation title cards, instead of being hidden behind a menu or requiring navigation away from their current page, they were more than twice as likely to use it \cite{ricks_does_2022}. 

While this study demonstrates the benefits of YouTube's user controls, there still exist challenges to its uptake to remove unwanted recommendations. First, we note that these controls could be used to create digital media environments that run counter to democratic norms of diversity and breadth of perspectives. Thus, policy makers should pay attention to the potential for user agency to further limit their capacity for and consumption of cross-cutting content.

Second, as our survey highlights, knowledge of these buttons is still an issue. Much of the general YouTube-using public was not aware that the ``Not interested" button exists, for example. Even more troubling was that even those who experienced unwanted recommendations recently -- thus having ample motivation to discover content removal tactics -- still had not become aware of the button. 

Third, user interaction flows from these buttons may violate design principles in a way that limits users' ability to fully understand and anticipate the effects of different user controls \cite{smith_dark_2021}. For instance, they found that users were not fully aware of their effects on recommendations and account settings, and so they shied away from using them at all. Further compounding users' hesitation to take up these features is the perception that some of their effects are irreversible. 

Lastly, the actions that these user control buttons allow are responsive, rather than proactive. Users respond to a poor recommendation by eliminating it, rather than asking YouTube to tailor their recommendations before they see it. Thus, the worrisome effects of misinformation, toxicity, and offensiveness may have already taken its harmful course by the time the user decided to eliminate them. Therefore, these features cannot be seen as a substitute for diligent and thorough content moderation by the platform. 
% ====================================================

% ====================================================
\section{Limitations}
\label{sec:limitations}

Our findings add to a growing chorus of studies investigating problematic and unwanted recommendations on YouTube. However, our study is not without its limitations. First, we perform channel-level, rather than video-level labeling. Performing labeling at this level may result in labeling channels as a certain topic even if not all of its videos are of that topic (e.g. an \emph{Alt-Right} channel sometimes posting music videos), or, conversely, labeling a channel as off-topic even if just a few of its videos are topical (e.g. a science vlogger who occasionally talks about their journey to atheism). However, we are encouraged by the fact that other studies have taken this approach \cite{tomlein_audit_2021,chen_exposure_2022}. Analyzing channels as a whole is still important because they indicate a high number of videos of that topic, and users may be encouraged to subscribe to these channels even if not all videos in the channel are topical.

Another limitation is that of generalizing from the particular settings our sock puppets used. In particular, our sock puppets tested just four topics, using geolocation of the US East AWS center in August of 2021. While we found that certain strategies work in these conditions to remove recommendations, we cannot be sure it does in other conditions. New topics may be harder or easier to scrub due to more or less general interest in that topic in the broader YouTube ecosystem, respectively. Our code is publicly available\footnote{\url{https://github.com/avliu-um/youtube-disinterest}} so that we and other researchers may continue to understand more general effects of our scrubbing strategies. 
% ====================================================

% ====================================================
\section{Conclusion}
\label{sec:conclusion}

With these results we conclude that different strategies to remove unwanted content on the YouTube platform work to different degrees, and from our tested strategies we found that using the ``Not interested" button was a clear winner. However, this strategy has not seen widespread adoption among users. That is, while those who know about these effective buttons get to experience their effective behavior, 44\% of adult YouTube users in the US are not aware that they exist. Thus, we join existing calls for YouTube to amplify more broadly the effective ways to remove unwanted recommendations on its platform.
% ====================================================

% ====================================================
\section*{Ethical Statement}
\label{sec:ethics}

We now discuss the ethical concerns of our study. First, since our sock puppets are computer scripted, we do not risk making real users watch potentially harmful content, such as those from the \emph{Alt-Right} channels. However, making the bots watch a lot of content from a given topic may still increase its prevalence on YouTube by boosting its general popularity. Also, pressing the ``Dislike" button on channels may cause them to be demoted in recommendations, limiting YouTube creators' ability to generate advertising revenue. 

While this is a possibility, we do not find these costs to outweigh the benefits of our study. First, we consider the potential cost to content creators of negative interactions with the system, such as pressing the ``Dislike'', ``Not interested'', ``Don't recommend channel'', and ``Delete from watch history'' buttons. Here we note that (a) our bots collectively injected up to 3 such interactions per video, which we expect is small compared to the number of ``authentic" ones, (b) we cleared all the revertible actions caused by the audit when it exited its experimental runs, and (c) the average lifetime of an audit in this study is less than six hours, including both the stain phase and scrub phase. This means that not only is the effect of negative interactions small per video, it is also both short-lived and fully reversed.

Another potential cost of our study is that our audits would irreversibly alter two public metrics -- total view count and total watch time. However, the costs are small because we do not expect to affect videos' and/or channels' overall prevalence by much because the number of views we are ``artificially" introducing to the YouTube world are minuscule in relation to the amount of ``authentic" views that the videos have received. 

Our study's findings are a benefit to all YouTube users alike, because they can inform users on how best to deal with and get rid of unwanted recommendations. We think these benefits outweigh the minimal harms.

As for the survey participants, we must make sure that they are not being put in the way of any harm. Because we did not ask them to view any actual content, but rather recall times in their life in which they had interacted with YouTube, we are only at the risk of having participants revisit potentially-triggering or traumatic events if they have had any. However, we include at our introduction of the survey a description of the survey, which describes the questions we wanted to ask them. Thus, if a participant expected such harms to occur to them, they would have not consented to the survey, and they would have been removed from the panel before seeing any questions.
% ====================================================

% ====================================================
\appendix

\section{Annotation Codebook}
\label{app:codebook}

\subsection{Alt-Right}

\subsubsection{Process}
\begin{itemize}
    \item ``Carefully inspect each one of the channels in this table, taking a look at the most popular videos, and watching, altogether, at least 5 minutes of content from that channel.'' -- \citet{ribeiro_auditing_2020}
    \item Search around for former connections to Alt-Right, e.g., Google them and then cross-reference with \citet{marantz_alt-right_2017}.
    \item Can also check existing lists of alt-right channels \cite{chen_exposure_2022,ribeiro_auditing_2020} and online forum.\footnote{\url{https://www.reddit.com/r/Fuckthealtright/}}
\end{itemize}

\subsubsection{Distinguishing the Alt-Right from within the Alternative Influence Network (AIN)}
\begin{itemize}
    \item Read \citet{noauthor_alt_2019} and \citet{marantz_alt-right_2017}, and \cite{weiss_opinion_2018}.
    \item The Alt-Right is the most extreme of three groups of the AIN \cite{lewis_alternative_2018}. The other two groups are the Alt-Lite (less extreme) and Intellectual Dark Web (least extreme).
    \item To distinguish between the Alt-Right and Alt-Lite, note that they are similar in their hatred of feminists, immigrants, social justice warriors, Jewish people, and others, but at the same time the Alt-Lite purport to reject white supremacist thinking that the Alt-Right espouses~\cite{marantz_alt-right_2017}.
\end{itemize}

\subsection{Antitheism}

\subsubsection{Process}
\begin{itemize}
    \item Watch intro video (if exists) and several other videos
    \item Look for anti-religion videos that could indicate potential antitheism
    \item Look for discussion of the channel/personality's own ideology -- do they talk about being atheist?
\end{itemize}

\subsubsection{Definition}
\begin{itemize}
    \item ``The (a) self-identified atheist who is also (b) actively critical of religion. Also called New Atheists or Street Epistemologists." -- \citet{ledwich_algorithmic_2019}
    \item The keyword here is ``self-identified". They should state their own religious beliefs somewhere in their videos. You can do some web searching to confirm.
    \item Some channels are scientific explanations of phenomena that may be traditionally explained by religion (for instance, explaining ancient floods with scientific reasoning). These videos should not be evidence for antitheism.
    \item Some channels are anti-religion, but only for certain religions (e.g., ex-Jehova's Witness members). However, if they do not talk about their own religious beliefs as being atheist, then it should not count as antitheism.
\end{itemize}

\subsection{Politics}

There are two types of requirements to satisfy here: (1) the format of the content and (2) the content itself. This codebook applies to both the politics-left and right topics.

\subsubsection{Decide whether the channel is ``political''}
\begin{itemize}
    \item Must cover US news and/or some sort of current events
    \item Must be in English
    \item Channels reposting clips of others can count as political
    \item Can be satire as long as the target’s political identity is clear and consistent
    \item If a channel only talks about religious or biblical content, then it is not political
\end{itemize}

\subsubsection{If the channel is political, then decide its political leaning}
\begin{itemize}
    \item Use the MBFC definition of left and right-leaning\footnote{\url{https://mediabiasfactcheck.com/left-vs-right-bias-how-we-rate-the-bias-of-media-sources/}}
    \item Can check both MBFC and All-Sides for leaning labels
\end{itemize}

\subsubsection{Special case - local news outlets}
\begin{itemize}
    \item First, check MBFC and All-Sides for label
    \item If that station does not have a label, then use that of the parent company
\end{itemize}

\subsubsection{Special case - late night shows}
\begin{itemize}
    \item If a show primarily makes fun of politicians, regardless of their affiliated political parties, then it is not political
    \item If a show consistently demotes politicians from one side, or consistently promotes issues fitted into one side’s agenda, it may be political-left or political-right
\end{itemize}
% ====================================================

% ====================================================
% AAAI reference format
\bibliography{main}
% ====================================================

\end{document}